# Improvised Broadcast Algorithm for Wireless Networks

Ashima Goel
Dept. of Computer Science and Engineering
NIIT University
Neemrana, India
ashima.goel@st.niituniversity.in

Debasis Das
Dept. of Computer Science and Engineering
NIIT University
Neemrana, India
debasis.das@niituniversity.in

*Abstract*—**Broadcasting problem is an important issue in the wireless networks, especially in dynamic wireless networks. In dynamic wireless networks the node density and mobility is high, due to which several problems arise during broadcasting. Two major problems faced are namely, Broadcast Storm Problem and Disconnected network problem. In a highly dense network, if information is being flooded in a loop, it could lead to broadcast storm. The broadcast storm may eventually crash the entire network and lead to loss of information. Mobility of the nodes may lead to the problem of Disconnected Network. If the two nodes sending and receiving information are mobile with different speeds, it could lead to a disconnection between them as soon as the receiver moves out of the communication range. In this paper, we are trying to solve both the problems based on our proposed algorithms.**

## I. Introduction

In a wireless network, there are no physical wired connections between communicating nodes.[2][3] Communication between nodes happens by using radio waves signal frequency. Broadcasting is a method of transmitting packets which are to be received by every device connected on the network.[1] When a new device is connected on the network, broadcast address is used to declare the addition of the device to all the other existing devices. Broadcasting resolves many issues and problems in network, especially in a mobile wireless network.[2][3][5]

In a mobile wireless network, the host nodes are also mobile. Along with communicating, transferring and receiving information, the nodes are dynamic in nature. It is not always possible to send data packets in a single hop to the destination node, since the host node is also mobile. Instead, data packets are sent in several hops by passing many immediate host nodes to reach the destination node. MANETs can be used for sending emergency warnings in battlefields, information about natural disasters, and sending information in scenarios where the locations of both the sender and receiver are dynamic.

Direct broadcasting of information among nodes in a wireless network can result in data redundancy and data collision especially in conditions where two or more nodes are sending information to the same node at the same time. This problem of data collision is termed as Broadcasting Storm[8]. Since in a mobile Wireless network, the nodes are mobile, hence the data has to be transmitted more quickly and repeatedly. Mobility can lead to sparse network situation where there is a huge probability that during the relaying of the message, receiver node might be out of the transmission range. If the speed of both the sender and receiver is same and are moving in the same direction/within the communication range, there is no issue. But if the speed of both the nodes is different, it could lead to disconnected networks. The node could move out of the range of sender node. In this situation, sending of data packets becomes a critical problem.

Most of the research work either takes care of the Broadcast storm problem or the disconnected network problem. In this paper, we propose an algorithm to effectively minimize and take care of the two problems.

## II. Priliminary And Related Work

There are many different techniques and protocols designed to solve the problem of Broadcast Storm and Disconnected Network. In VANETs, Distributed Vehicular BroadCAST (DV-CAST) is used.[6] It uses only the information about the local topology for handling broadcast message. Slotted p-persistence broadcast rule is also one example, where the node verifies the ID of the incoming packet and decides whether to accept the packet or not. Some protocols also use the information from the GPS to prevent broadcast storm. In the sparsely connected network, different algorithms are used.[10] Delay Tolerant Network for Disconnected Networks is such example.[4][9][1] It is used in real like situations where there may not be any defined end to end route between the nodes. Although efficient techniques have been designed to mitigate these problems, there are only few techniques which consider both the sparsely dense network and highly dense network together in a distributed network.

## III. Problem Statement

In real world scenario, the major issues revolving around the broadcasting in dynamic wireless network are data collision, data redundancy and disconnectivity. A densely populated network can lead to choking of the transmission medium. The probability of two or more neighbors broadcasting message to the same node increases. Second problem that arises is when the network is sparsely populated. In a sparse network, there is a possibility of the nodes getting disconnected. The problem statement is to find a solution which helps in efficient transmission of data from one node to another in a densely populated network and to find connected nodes in a sparsely populated network to transmit data packets to the destination nodes.

## IV. PROPOSED BROADCAST ALGORITHM

Through the proposed algorithm, we solve the problems identified in the earlier section. Dynamic wireless network face two major problems, disconnected network and Broadcast Storm. Our proposed algorithm solves the issue of data collision, redundancy and disconnected network through Algorithm for Disconnected Networks and Algorithm for Broadcast Storm. The objective of the algorithm for Broadcasting is to identify which problem the transmission is facing and use the other two algorithms to solve them.

### A. Algorithm for Broadcasting

The below algorithm is our main code, where it identifies whether the transmission and network face any of the two above stated problems. Solving the problems, if any, it transfers the data packet to the destination node, displaying the message of successful transmission.

#### 1) Data Structures Used

**int connected( i , j )** returns 0 if two nodes i and j are not connected and 1 if nodes i and j are connected

**int datacollision( i , j )** returns 1 when any node is sending data to node j except node i at the same time and 0 when none are sending data at the same time to node j

**int transfer_data_packet( i , j )** will transfer data packet from node i to node j and return 1 if transfer is successful and 0 if transfer fails.

**int finding_neighbour( i, j )** returns the nearest neighbor to node i on the path s to d where the new node returned is not j and -1 if no neighbor exists.

**int solve_disc_network( i , j )** returns a new node to transfer data such that data eventually reaches 'd'. It returns -1 if no node is present.

**int storm_broadcasting( j )** returns 1 if the collision is resolved and 0 if not.

**display()** function prints the final status of transmission between two nodes

#### 2) Steps for Broadcasting

Let s be the source node and d be the destination node in the Dynamic Wireless Network
Let R be the communication range
Let N be the entire set of nodes in a communication range
Let i and j be the ith and jth node in the network where
$i \in N$ and $j \in N$
s is the source node and d is the destination node.
int flag indicate successful transfer if flag = 1 and transfer fail when flag = 0

Step 1:  Initialize a node,  i = s;
     j= finding_neighbour( i, i );
     int k =0;
Step 2:  while ( i != d && j >= -1)
   if ( connected( i , j ) == 0 ) then
     j = solve_disc_network( i , j );
    if ( j > 0 ) then
     if(collision( i , j ) == 1) then
  {
  k = storm_broadcasting( j )
  flag = 1;
  go to Step 5
    }
Step 3:  flag = transfer_data_packet(i,j);
Step 4:  if( flag == 0 ) then
    {
  display( "error" );
  go to Step 7
     end;
  }
Step 5:  display (success)
Step 6:  i = j;
   j = finding_neighbour( i , i);
   end of while;
Step 7:  if( i == d && j ! = - 1) then
    display("Message successfully sent!") ;
  else
    display("error in transmission");
  end;

#### 3) Explanation

**Step 1:** In the above mention algorithm, we set any node i to be equal to to node s (source node) and j to be any neighbor of i lying in the path s to d (source to destination).

**Step 2:** Inside while loop we have put the terminating condition where either node i = d (reached destination node) or j = - 1 (no other node is there in the path to reach destination node). Inside the loop we check if two nodes i and j are connected or not (using connected( i , j )). If the two nodes are not connected( function returns 0) then we find a new node j (using solve_disc_network( i , j )) where i and j are connected or returns -1 if no other node exists in the path. Once nodes i and j are connected we check for collision. If a collision exists, we go inside the if condition, and calls storm_broadcasting( j) function where node j selects from which node to accept data transfer from. It returns 1 if the transfer is successful and jumps to Step 5. If there is no collision we go to Step 3.

**Step 3:** transfer of data from node i to j (transfer_data_packet( i , j )). The flag variable indicated successful transfer of data packet from i to j.

**Step 4:** If flag value is 0 i.e the data has not been transferred we display the error message and go to Step 7

**Step 5:** Display success if flag = 1

**Step 6:** Now make node i equal to j and find a new j (neighbor of new i) and continue the loop till terminating condition is met

**Step 7:** If the node i and destination node d becomes same and k variable is node is positive no. display message of successful transfer, otherwise display failure in transfer.

*B. Algorithm for Disconnected Newtork Problem*

In this algorithm we solve the disconnected network problem. The problem arises when a data needs to be broadcasted to nodes and there exists disconnectivity among them. The function returns a new connected node in the network such that the new node lies in the path of source node and destination node.

*1) Data Strucutres Used*

**find_previous_node(prev)** returns the parent node of the prev node

*2) Steps to solve Disconnected Network Problem*

If the counter 'c' goes more than 2(or any limit set by the admin) then sending data become redundant. Hence return -1.

Step 1:  int solve_disc_network(node I, node j){
  j = finding_neighbour(i , j);
Step 2:  return j ;
  }

*3) Explanation*

This function is very simple. We take in parameter i and j.
**Step 1:** We call the function finding_neighbour(i, j) to find a new neighbor to i such that the new neighbor is not node j. Here j could be the new node or -1 if no other node exists.
**Step 2:** Then we return j.

*C. Algorithm for Broadcasting Storm Problem*

Broadcast storm problem occurs when the network is highly dense. There can be multiple nodes trying to broadcast the information to a single node at the same time which leads to problems like data redundancy and collision. With Algorithm 3, we have use the approach of probability, where the pre-defined probability will decide from which node we should accept data.

*1) Data Structures Used*

**node_max_prob(i)** returns the neighbor node of 'I' with highest probability

*2) Steps to solve Broadcasting Storm Problem*

Let two or more nodes are sending data packets to node 'j' at the same time.

  int storm_broadcasting( int j )  {
Step 1:  Initialize variable, flag2 = 0;
Step 2:  node max =  node_max_prob(i)
Step 3:  flag2 = transferdatapacket(max, j)
Step 4:  Put rest of the nodes on waiting till the data packet of 'max' node is transferred successfully
Step 5:  return flag2;

*3) Explanation*

We take a integer node as a parameter to the function storm_broadcasting.
**Step 1:** Initialize variable flag2 = 0,which would indicate whether the data has been transferred to the j from the selected node.
**Step 2:** node j reads the probability of all the neighbor nodes and finds the node with maximum probability using the function node_max_prob(i) which returns the node with highest probability.
**Step 3:** We call the transfer_data_packet() function to complete the data transfer
**Step 4:** All the rest of the nodes are put on wait till the transfer is made.
**Step 5:** Return flag2 to indicate whether the transfer was successful or not.

## V. CONCLUSION

In this paper, both the problem of storm collision and disconnected networks is successfully taken care by the above techniques. Dynamic Wireless networks are deployed in situations of utmost importance such as emergency warning, war time machines etc. In dense networks, storm broadcasting techniques are used to help mitigate the problem of data collision and lose. Where as in a sparse network, new node connections are established to transfer the data . Hence by the above proposed algorithms we have successfully taken care of the commonly acknowledged problems in dynamic wireless network.

## VI. FUTURE SCOPE

In future, we will implement the proposed algorithm in the NS-2 simulator. Here in our research we assumed the network to be mesh network. The research can be further extended for other types on network topologies.